\documentclass[prb,twocolumn]{revtex4}
\usepackage{graphicx}

\newcommand{\beq}{\begin{equation}}
\newcommand{\eeq}{\end{equation}}
\newcommand{\beqa}{\begin{eqnarray}}
\newcommand{\eeqa}{\end{eqnarray}}

\begin{document}

\title{Transport through normal metal -- graphene contacts}

\author{Ya.~M.~Blanter}
\affiliation{Kavli Institute of Nanoscience, Delft University of
Technology, Lorentzweg 1, 2628 CJ Delft, The Netherlands}

\author{Ivar Martin}
\affiliation{Theoretical Division, Los Alamos National Laboratory, Los Alamos,
New Mexico, 87544, USA}

\date{\today}
\begin{abstract}

Conductance of zigzag interfaces between graphene sheet and normal metal is
investigated in the tight-binding approximation. Boundary conditions, valid for
a variety of scattering problems, are constructed and applied to the normal
metal -- graphene -- normal metal (NGN) junctions.  At the Dirac point, the
conductance is determined solely by the evanescent modes and is
inversely proportional to the length of the junction. It is also
independent on the interface resistance. Away from the Dirac point,
the propagating modes' contribution dominates.  We also observe that
even in the junctions with high interface resistance, for certain
modes, ideal transmission  is possible via Fabry-Perot like resonances.
\end{abstract}
\maketitle

\section{Introduction}

Recent experimental studies \cite{Geim,Kim,deHeer,Morpurgo} uncovered unusual
properties of graphene (graphite monolayers and bilayers), strongly contrasting
with the common knowledge inherited from studies of metals. This difference
originates from the fact that electrons in graphene monolayers obey the Dirac
(rather than the Schr\"odinger) equation. Thus, one has an opportunity to study
properties of the Dirac fermions in a table-top experiment. Predictions of
relativistic effects including the Klein tunneling \cite{Klein} and
Zitterbewegung \cite{Trauzettel} have been made. ``Ordinary'' phenomena, {\em
e.g.}, quantum Hall effect \cite{QHE}, weak localization \cite{Falko}
or Andreev reflection \cite{Beenakker}, are also strongly modified in
graphene as compared to normal metals. 

The most easily accessible measurements in graphene are those of
electrical transport.  Theoretically, one way of understanding
them is to extend the Landauer theory to graphene sheets,
considering them as a junction between two reservoirs. So far, a
common point was to describe reservoirs as bulk disordered
graphene \cite{Beenakker,Guinea,Tworzydlo}. This approach
considerably facilitates calculations; however, its relation to
the experimental situation, with contacts made of normal metals,
requires additional clarification. Indeed, the bulk in the available
graphene devices is defect free \cite{ripples}. Thus, the major
source of electron scattering in graphene are the boundaries, in
particular, contacts, and even qualitative understanding of
electric transport can not be achieved without careful
consideration of  electron behavior in the contacts.

Experimentally, graphene flakes are contacted by tunnel junctions
located on top of the flakes. Conceptually, we can discriminate
between three types of the junctions. One situation is when tunneling
from the normal 
reservoir to graphene occurs just at one point --- for instance,
due to the non-uniform thickness of the oxide layer in the contact
in combination with the exponential dependence of the tunneling
amplitude of this thickness. Obviously, in this case the voltage
between normal reservoir and graphene drops at the junction, and
the chemical potential in graphene can be considered as fixed. For
this point-like tunneling the resistance of a NGN junction depends
on the distance between the two tunneling points.

\begin{figure}[h]
\includegraphics[width=.7\columnwidth]{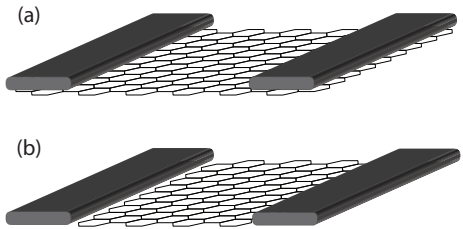}
\caption[]{(a) Typical experimental contact configuration: Metallic
  leads overlaying graphene.  The contact occurs over many microlinks
  which leads to equilibration of the chemical potentials in the metal
  and in graphene.  In the case of tight coupling under the contact
  the band structures of graphene  and  metal are expected  to
  ``fuse". (b) Theoretical model of a contact -- tunneling between the
  metal leads and graphene zig-zag edges.  It can be thought of as a
  limiting case of strong coupling (a) with the tunable  hybridization
  between contact and the edge. 
\label{fig:contacts} }
\end{figure}

More interesting, and apparently more experimentally relevant is
the situation when tunneling occurs at many points, covering a
large area (Fig. \ref{fig:contacts}). In this case, there is no
voltage drop on the contact, {\em i.e.} it becomes Ohmic.  In this
area under the contact, 
the wave functions of electrons from the normal reservoir and
graphene quasiparticles hybridize, forming a new substance, a
hybrid between graphene and normal metal. The nature of this
substance depend on the exact properties of the contacts, however,
it is reasonable to assume that generally it will be either closer
to graphene, or to a normal metal. In the former case, effective
description of a NGN system as a graphene -- graphene -- graphene
(GGG) contact, where the voltage is applied to the graphene, is
appropriate. In the latter case, the system is essentially a {\em
planar} NGN junction, where the voltage is applied to the normal
metal and drops at the interface between the normal metal and
graphene. It is also clear that properties of such contacts will
considerably depend on the relation between lattice periods of
graphene and normal metal, for instance, on whether the
lattices are commensurate or not.

Below, we discuss the latter type of junctions ---
planar NGN junctions, to complement the earlier studies of GGG
junctions. A first step in this direction has been
recently made by Schomerus \cite{Schomerus} who compared
resistances of NGN contacts with a zigzag interface and GGG
contacts. He considered the special 
case of equal overlap integrals between all neighboring sites in
the tight-binding model. He found that if the graphene sheet is
biased to the Dirac point, so that there are no propagating modes
through graphene, the difference between NGN and GGG junctions is
only quantitative. Outside this regime, the behavior of the
contacts may strongly depend on the type of the leads and the
interface. Below, we study this dependence for the simplest case
of square-lattice leads and zigzag interfaces, and find a number
of interesting results.

In this Article, we consider within the tight-binding
approximation the general problem of transmission through the NG
interfaces and NGN structures for arbitrary overlap integrals
$t_s$, $t_g$, and $t'$, in square-lattice normal metal, graphene,
and at the interface, respectively.  From the lattice
Schr\"odinger equation, we derive the wave-function matching
conditions at the interfaces. We apply them first to the
scattering off the NG interface and determine the reflection
probability back to the normal metal.

Then, we turn to the conductance of an NGN junction.  It is the
sum of two contributions originating from propagating and evanescent 
modes in graphene.  The number of the available propagating modes
is proportional to the radius of the Dirac cone $q_G$ taken at the
gate voltage energy $eV_G$, $g_G\propto V_G$.  Their contribution
to conductivity is therefore proportional to $V_G$; it is
independent of the graphene length $L$ in the limiting cases $q_GL
\ll 1$ and $q_GL \gg 1$.  At the Dirac point ($V_G = 0 $), there
are no propagating modes; however, there is a contribution to
conductivity from the evanescent modes that obeys the Ohm's law,
{\em i.e.} is inversely proportional to the length of the graphene
strip $L$ \cite{Tworzydlo,Schomerus}.  For non-zero gate voltage
$V_G$ the contribution of evanescent modes crosses over to
$L^{-1}(q_GL)^{-3}$ behavior for $q_GL \gg 1$.  We also observe
that for the symmetric case $t'^2 = t_st_g$, a narrow graphene
strip is ideally transmitting, both through the propagating and
evanescent states. Moreover, the conductance of such strip precisely
at the Dirac point is intependent of the interface overlap integral
$t'$. Even for opaque interfaces, $t'^2 \ll
t_s t_g$, we find that some modes transmit ideally, in a direct
analogy to the Fabry-Perot resonances in a double-barrier
structure.

\section{NG interface}

\begin{figure}[h]
\includegraphics[width=.7\columnwidth]{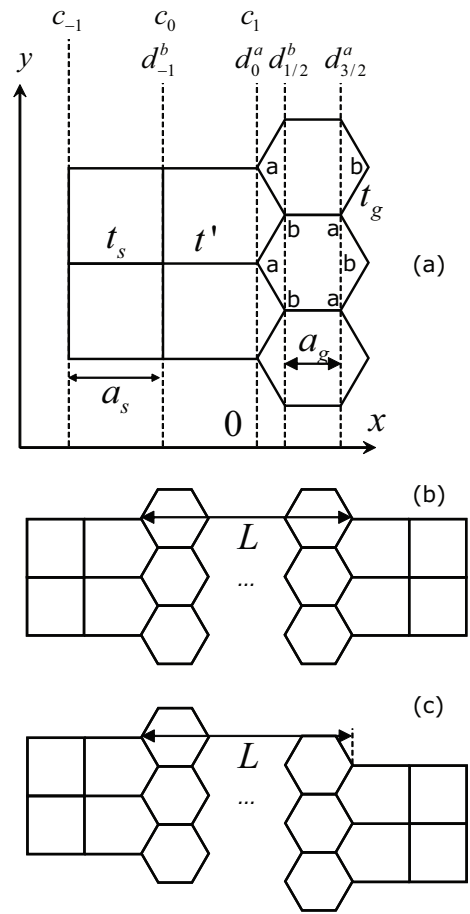}
\caption[]{(a) A zigzag interface between a square and graphene
lattices. Sites belonging to the two sublattices in graphene are
shown by letters $a$ and $b$. On the top, boundary conditions are
sketched: We introduced two columns of fictitious amplitudes,
$c_1$ in graphene, and $d^b_{-1}$ in the square lattice, used for
wave functions matching at the interface. (b, c) A NGN contact of the
width $L = a_g (3N +1)/2$ with an odd-integer (b) or an even-integer
(c) value of $N$.
\label{fig:zigzag4} }
\end{figure}

We consider a zigzag interface between normal metal with the
square lattice and graphene with the bond lengths $a_s$ and $a_g$,
matched such that $a_g = a_s/\sqrt{3}$, respectively,
Fig.~\ref{fig:zigzag4}. We assume 
that the normal metal band is taken near the half-filling, so that
the Fermi surface is nearly a square, whereas the graphene is
tuned close to the Dirac point. In the normal metal, the wave
functions are plane waves, $c (\mbox{\bf \em r}) = \left( e^{ik_sx}
+ re^{-ik_sx} \right) e^{ik_yy} \ , x < 0 $ , defined on the sites
of the square lattice. First and second terms describe incident
and reflected waves, respectively. In graphene, one has to define
two transmitted waves corresponding to two sublattices $a$ and
$b$, $[d^a(\mbox{\bf \em r}), d^b(\mbox{\bf \em r})]$.

For the zigzag interface, the wave vector component along the interface $k_y$
is conserved, and thus the two-dimensional scattering problem reduces to a
collection of one-dimensional problems for different values of $k_y$.  The
transverse component of wavevector ($k_x$) is not conserved and must be found
from the energy conservation.

In the tight-binding model, the amplitudes obey the lattice
Schr\"odinger equation. In particular, at the interface, we
obtain the following set of equations,
\begin{eqnarray}\label{contdiscrete2}
E c_0 & = & -2t_s c_0 \cos k_ya_s - t_s c_{-1}  - t'd^a_0 \ ; \nonumber \\
Ed^a_0 & = & -t' c_0 -2 t_g\cos
\frac{\sqrt{3}k_ya_g}{2} d^b_{1/2} \ ; \nonumber \\
Ed^b_{1/2} & = & -t_g d^a_{3/2} -  2 t_g \cos
\frac{\sqrt{3}k_ya_g}{2} d^a_{0} \ ,
\end{eqnarray}
where we moved to discrete notations, explained in
Fig.~\ref{fig:zigzag4}: The subscript shows the distance from the
interface, measured in corresponding lattice constants.

Away from the interface, the solution of the Schr\"odinger
equation is a superposition of plane (or evanescent, see below)
waves, with the energies
\begin{eqnarray} \label{ensquare}
E_s & = & -2t_s \left( \cos k_s a_s + \cos k_y a_s \right) \ , \\
E_g & = &\pm t_g \sqrt{1 + 4 \cos \frac{\sqrt{3}k_ya_g}{2}\left(
\cos \frac{\sqrt{3}k_ya_g}{2} + \cos \frac{3k_ga_g}{2}\right)} \ .
\nonumber
\end{eqnarray}
In this Article, we are mainly interested in the low-energy
regime, $E \ll t_g$. In this regime, the energy can be expanded in
the vicinity of the points where it turns to zero (the Dirac
K-points), $k_g 
= q_x$, $k_y = \pm 4\pi/(3\sqrt{3}a_g) + q_y$, $E_\pm = \pm3t_g\,
q/(2a_g)$, with $q = (q_x^2 + q_y^2)^{1/2}$.  For concreteness, we
only consider the ``upper" Dirac point, $k_y \approx
4\pi/(3\sqrt{3}a_g)$; by symmetry, the ``lower" one contributes to
transport identically.  For propagating waves, from the graphene
(lattice) Schr\"odinger equation we obtain $d^b = i e^{i\varphi}
d^a$, where $\tan\varphi = q_y/q_x$.

The interface equations, Eqs.~(\ref{contdiscrete2}) can be
equivalently recast
in the form of the wave function matching, which is convenient when
considering
more complex scattering, such as NGN (see below). This can be done by
continuing the wave functions across the interface. Introducing
fictitious wave
function elements $c_1$ and $d_{-1}^{b}$, the first two equations of
Eqs.~(\ref{contdiscrete2}) become
\beqa
t' d_0^{a} = t_s c_{1} \, ;\  t_g d_{-1}^{b} = t' c_0 \ .\nonumber
\eeqa
These are general boundary conditions and can be applied to a scattering
problem with arbitrary arrangements. For example, scattering from N to
G can be obtained if we use
\begin{displaymath}
c_0 = 1 + r \ ; c_1 = e^{ik_s a_s} + r e^{-ik_s a_s} \ ;
d_{-1}^{b} = d_{0}^{b} e^{-i k_g a_g} \ .
\end{displaymath}
Substituting into the above boundary conditions, we find the
equations connecting the amplitudes of the waves,
\beqa
t_s (e^{ik_s a_s} + r e^{-ik_s a_s}) = t' d_0^{a}\ ;\\
t' (1+r) = t_g d_{0}^{b} e^{-i k_g a_g}  \ .
\eeqa
Thus, for the reflection coefficient $R = \vert r \vert^2$, at low energies
($k_g a_g \ll 1$), we find
\begin{equation} \label{refl}
R = \frac{\beta + \beta^{-1} - 2 \sin ( \varphi + k_sa_s)}{\beta + \beta^{-1} -
2 \sin (\varphi - k_sa_s)} \ , \ \ \beta \equiv \frac{t'^2}{t_st_g} \ .
\end{equation}
Near half-filling, $k_sa_s \approx \pi/3$;
however, these results apply to the case of arbitrary filling, where $k_sa_s
\ne \pi/3$.  As anticipated, for an opaque interface $t'^2 \ll t_at_s$, the
reflection approaches 1. Note also that one can only match the wave vectors in
the $y$-direction for a very limited set of wave vectors around $k_sa_s =
\pi/3$; all other states are ideally reflected by the interface.

\section{NGN contact: propagating and evanescent modes}

Consider now a graphene sheet of length $L$ connected to two
square-lattice metal electrodes (with the same overlap integral $t_s$) at
two ideal zigzag interfaces. Note that
such an arrangement is only possible provided $L = a_g (3N + 1)/2$,
with an integer $N$. For an odd-integer $N$ the square lattice leads
are aligned (Fig. \ref{fig:zigzag4}b), for an even-integer $N$ they
are shifted by half a period (Fig. \ref{fig:zigzag4}c). Note that
sites at the two interfaces always belong to different graphene
sublattices. We describe both cases on equal footing.

We take the wave function in the left
electrode in the same form as before, $c (\mbox{\bf \em r}) =
\left( e^{ik_sx} + re^{-ik_sx} \right) e^{ik_yy}$, the wave
function in graphene as a combination of left- and right-moving
waves for each sublattice, and the wave function in the right
electrode as the transmitted wave $f (\mbox{\bf \em r}) =
w\exp(ik_s(x-L) + ik_y y)$.

The equations for the NGN structure read
\beqa
t' c_0 &=& t_g (d_l^{b} + d_r^{b})_{-1} \ ;\\
t_s c_1 &=& t' (d_l^{a} + d_r^{a})_0 \ ;\\
t' f_0 &=& t_g (d_l^{a} + d_r^{a})_{(3N+1)/2} \ ;\\
t_s f_{-1} &=& t' (d_l^{b} + d_r^{b})_{(3N+3)/2} .
\eeqa
In addition to $c_1$ and $d^b_{-1}$, we introduced two more ``unphysical"
amplitudes, $d^{a}_{(3N+3)/2}$ and $f_{-1}$.  We use now $d_r^{b} = -i e^{i
\varphi} d_r^{a}$ and $d_l^{b} = i e^{-i \varphi} d_l^{a}$, where $\varphi$ in
both expressions is defined for the right-moving electrons. Solving this
scattering problem for $q a_g \ll 1$, we find for the transmission amplitude
\begin{eqnarray} \label{transmampl1}
w & = & -2i \cos \varphi  \sin k_sa_s \,e^{ik_sa_s} J^{-1}\ ,\\
J & = & \beta^{-1} e^{-ik_sa_s} \cos (\varphi - qL \cos\varphi) - 2\sin
(qL\cos\varphi) \nonumber \\
& - & \beta e^{ik_sa_s} \cos (\varphi + qL \cos \varphi).
\nonumber
\end{eqnarray}
Note that for $qL \ll 1$ and $t'^2 = t_s t_g$, the junction is ideally
transmitting.

For a finite length of the graphene strip, there are also
solutions which are not propagating, but rather exponentially
increasing or decreasing as $e^{\pm\eta_x x}$ -- the {\em
evanescent states}.  Their energy reads
$$E_{\pm} = \pm \frac{3t_g}{2} \sqrt{q_y^2-\eta_x^2},$$
which is defined as far as $|\eta_x| \le |q_y|$. Similar to the
propagating case, we can find the relation between the components
of the graphene wavefunction,
\beq\label{eq:dadb}
d^{b} = \pm \sqrt{\frac{q_y + \eta_x}{q_y - \eta_x}} d^{a},
\eeq
where $\pm$ correspond to $E_\pm$.  In the following, we consider
the positive energy branch, $E_+$.  While there is no propagation
in the evanescent case, we can still define the right ($\eta_x
>0$) and left ($\eta_x < 0$) components of the wave function
-- the wave ``propagates" in the
direction of decay.  For these components, $d_r^{b} = Z d_r^{a},\
d_l^{b} = Z^{-1} d_l^{a}$, where $Z=\sqrt{({q_y + |\eta_x|})/({q_y
- |\eta_x|})}$. The transmission amplitude through the evanescent
modes becomes
\begin{eqnarray}\label{eq:we}
w & = & {-2i \sinh \zeta\,\sin{k_s a_s}\,e^{ik_sa_s}}{\tilde
J}^{-1} \ , \label{w_evan}\\
\tilde J &=& \beta^{-1}e^{-ik_s a_s}\,\sinh (\eta_x
L +\zeta) - 2 \, \sinh \eta_x L \nonumber \\
&+& \beta e^{ik_s a_s}\,\sinh (\eta_x L -\zeta) \ , \nonumber
\end{eqnarray}
with $\zeta = \ln Z$. Clearly, for $t_g t_s = t'^2$ and $\eta_x L
\ll 1$ we obtain again the perfect transmission.

\section{Current and conductance}

The current through the junction is expressed as
\begin{eqnarray} \label{curgen1}
I_x & = & eW \int \frac{dk_sdk_y}{(2\pi)^2} v_x \left\vert w(k_s,
k_y) \right\vert^2 \nonumber \\
& = & \frac{eW}{(2\pi)^2 \hbar} \int_{eV_G}^{eV_G + eV} dE \int
dk_y \left\vert w(k_s, k_y) \right\vert^2 \ ,
\end{eqnarray}
where $W$ is the width of the graphene strip in the $y$-direction, and $v_x =
\hbar^{-1}
\partial E_s/\partial k_s$ is the group velocity. This expression
includes the contribution from both propagating (real $k_s$) and
evanescent (purely imaginary $k_s$) states.

Let us first consider the contribution of {\em propagating
states}. The integration is carried over the momenta for which
$v_x > 0$. In the linear regime, from Eq. (\ref{curgen1}) we
obtain the conductance
\begin{equation} \label{condgen1}
\frac{G_{tr}}{G_Q} = \frac{Wq_G}{\pi} \int_{-\pi/2}^{\pi/2} \vert
w \vert^2 \cos \varphi d\varphi \ ,
\end{equation}
with $q_G \equiv {2e\vert V_G \vert}/({3t_ga_g})$ and $G_Q =
e^2/2\pi\hbar$ being the conductance quantum. We were able to
analyze Eq. (\ref{condgen1}) analytically in the two limiting
cases, short ($q_GL \ll 1$) and long ($q_GL \gg 1$) junctions.

For short junctions, $q_GL \ll 1$, the transmission coefficient,
\begin{displaymath}
\vert w \vert^2 = \frac{4\beta \sin^2
k_sa_s}{\beta^{-1} + \beta - 2  \cos k_sa_s} \ ,
\end{displaymath}
does not depend on the angle $\varphi$, and we obtain
\begin{equation} \label{transmshort}
\frac{G_{tr}}{G_Q} = \frac{2Wq_G}{\pi} \frac{4\beta \sin^2
k_sa_s}{\beta^{-1} + \beta - 2  \cos k_sa_s} \ .
\end{equation}
The quantity $2W q_G/\pi$ can be interpreted as a ``number of
transport channels." In the case $\beta = 1$, the conductance
equals $2Wq_GG_Q/\pi$.

For long junctions, $q_GL \gg 1$, we use the fact that
$\cos(q_GL\cos\varphi)$ is a rapidly oscillating function of the angle
$\varphi$. In particular, for $\beta = 1$, we have
\begin{eqnarray} \label{transmlongsym}
& & \vert w \vert^2 =  \cos^2 \varphi \sin^2 k_sa_s  \left[ \sin^2 k_sa_s
\cos^2 \varphi \cos^2 (q_GL\cos\varphi)
\right. \nonumber \\
& & + \left. (1 - \cos k_sa_a \sin \varphi)^2 \sin^2
(q_GL\cos\varphi)\right]^{-1} \ .
\end{eqnarray}
In this situation, the integral in Eq. (\ref{curgen1}) can be
discretized. Indeed, between the points $\varphi_n$ and
$\varphi_{n+1}$, such that $\cos \varphi_n = \pi n/(q_GL)$, the
integral in Eq. (\ref{curgen1}) can be easily calculated assuming
that the slow functions $\cos\varphi$ and $\sin\varphi$ are
constant and equal to $\cos\varphi_n$ and $\sin \varphi_n$
everywhere except for in combination $\cos(q_GL\cos\varphi)$. Then
Eq. (\ref{curgen1}) becomes a discrete sum over the periods $n$ of
the function $\cos(q_GL\cos\varphi)$. Converting the sum into an
integral (the integrand is a smooth function of $n$), one obtains
\begin{equation} \label{condlongsym}
\frac{G_{tr}}{G_Q} = \frac{Wq_G (1 - \sin k_sa_s)}{4 \sin k_sa_s
\cos^2 k_s a_s} \ .
\end{equation}
This result is length independent, similar to Eq.
(\ref{transmshort}), and for $k_s a_s = \pi/3$ the conductance of a
long graphene layer is suppressed as compared to a short layer.

For untransparent interfaces, $\beta \ll 1$, the transmission
coefficient can be approximated as
\begin{equation} \label{translongdirty}
\vert w \vert^2 = \frac{4 \beta^2 \sin^2 k_sa_s \cos^2
\varphi}{\cos^2 (q_GL\cos\varphi - \varphi)^2 + 4\beta^2\sin^2
k_sa_s \sin^2 (q_GL\cos\varphi)} \ .
\end{equation}
This expression has the structure similar to that of resonant
tunneling for a double barrier: Typically the numerator is of
order of $\beta^2 \ll 1$, whereas the denominator is of order $1$,
and the transmission probability is small. However, for certain
directions $\varphi_m$ of the wave vector, when the cosine in the
denominator vanishes, the transmission becomes ideal. One can
expand the expression around the resonance $\varphi_m$ to obtain
the Breit-Wigner structure of the resonance, $\varphi = \varphi_m +
\delta\varphi$, $\delta\varphi \ll 1$,
\begin{equation} \label{translongdirty1}
\vert w \vert^2 = \frac{4 \beta^2 \sin^2 k_sa_s \cos^2
\varphi_m}{(q_GL \sin\varphi_m - 1)^2 \delta\varphi^2 + 4\beta^2\sin^2
k_sa_s \cos^2
\varphi_m} \ .
\end{equation}
Typically one can omit $1$ as compared to $q_GL\sin\varphi_m$ in
the denominator. The main contribution to the current comes from
the directions around the resonances ($\varphi$ close to
$\varphi_m$). Integrating the Breit-Wigner expressions
(\ref{translongdirty1}) and transforming the resulting sum over
$m$ into an integral, we obtain
\begin{equation} \label{condlongdirty}
G_{tr}/G_Q = \beta Wq_G \sin k_sa_s \ .
\end{equation}
Note, that here the conductance of a long layer is proportional to $\beta$ and
thus parametrically exceeds the conductance of a short layer (proportional to
$\beta^2$, Eq. (\ref{transmshort})). This effect is due to the resonant
structure of Eq. (\ref{translongdirty}).

Let us turn now to the contribution of the {\em evanescent modes}.
In the vicinity of the Dirac point, the number of propagating
states vanishes proportionally to $E$, and thus the contribution
of the large number of evanescent states with $|\eta_x| \approx
|q_y|$ becomes dominant. At zero energy, $E\approx +0$, and $q_y
>0$, one has $Z \rightarrow \infty$. Thus Eq. (\ref{w_evan}) becomes
\beq\label{eq:w+}
w_+ = \frac{-2 i \sin k_s a_s \,e^{ik_s a_s}}{\beta^{-1}e^{-ik_s
a_s}\,e^{\eta_x L} - \beta e^{ik_s a_s}\,e^{-\eta_x L}}.
\eeq
Interestingly, for negative values of the deviation $q_y<0$ from
the $K$ Dirac point, the transmission amplitude for zero-energy
states ($E\approx +0$ and $Z \rightarrow 0$), is different:
\beq\label{eq:w-}
w_- = \frac{-2 i \sin k_s a_s \,e^{ik_s a_s}}{\beta^{-1}e^{-ik_s
a_s}\,e^{-\eta_x L} - \beta e^{ik_s a_s}\,e^{\eta_x L}}.
\eeq
Although for $\beta = 1$ the probabilities are the same,
$|w_-|^2=|w_+|^2$, for $\beta \ne 1$, they are different. This
effect is related to the chirality of the surface states in a
given Dirac valley. Indeed, a zigzag interface of graphene supports
a continuum of chiral edge states at zero energy \cite{dress}: They can only
propagate in the direction $q_y > 0$ for the lower Dirac point K and in the
direction $q_y > 0$ for the upper Dirac point. These states decay
exponentially in the bulk of graphene. In a graphene nanoribbon,
which is essentially our graphene sheet disconnected from the
electrodes, these states become evanescent modes \cite{fertig}. It is
this chiral nature of the surface states that cases an asymmetry
between the two directions of $q_y$. 

Note that the expression for $w_-$ can be obtained from $w_+$ by
flipping the sign of $\eta_x$; thus integration over positive and
negative values of $q_y$ in the conductivity is equivalent to
integration over positive and negative values of $\eta_x$.  The
tunneling probability is
\beq
|w|^2 = \frac{4 \sin^2{k_s a_s}}{\beta^{-2}e^{2\eta_xL} +
\beta^{2}e^{-2\eta_xL} -  2\cos 2k_s a_s} \ ,
\eeq
It becomes ideal for $\beta = \exp(\eta_x L)$. 

The conductance including now both Dirac points is
\beqa\label{Geva}
\frac{G_{ev}}{G_Q} &=&  \frac{W}{\pi L}
\int_{-\infty}^{+\infty}{d\eta_x}|w|^2\\
&=&  \frac{W(\pi - 2 k_s a_s)}{\pi L}\tan{k_s a_s} \ .
\eeqa
Surprisingly we find that the resulting conductivity is
independent of the value of $\beta$, even for small $\beta$ (weak
contact between the metal and graphene)!  By inspection one can see
that in this limit the major 
contribution comes from $w_-$, i.e. the negative values of $q_y$.
In the regime $\beta \ll 1$, the graphene region reduces to the
zig-zag nanoribbon, which is known \cite{fertig} to have a continuum
of surface states for $q_y < 0$ with energies that scale as
$e^{-|q_y|L}$. Precisely at the Dirac point these states only have
non-zero amplitudes at one of the sublattices. The conductivity is
dominated by the tunneling through these states. A similar effect was
discussed in Ref. \onlinecite{Nakanishi} for tunneling between a carbon
nanotube and a metallic electrode. 

Still, Eq.~(\ref{Geva})
clearly cannot hold for $\beta = 0$. To establish the limits of
its applicability we analyze the terms that were dropped while
going from Eq.~(\ref{eq:we}) to Eqs.~(\ref{eq:w+}) and
(\ref{eq:w-}). Taking the limit $\zeta \rightarrow \pm \infty$ is
only valid for exact zero-energy states. However, for finite
transport or gate voltage such that $q_G \ll 1/L$, $Z(q_y <
0)\approx q_G/|q_y|$.  The conductivity Eq.~(\ref{Geva}) is
dominated by terms with $|q_y| \approx |\eta_x| \sim 1/L$.  The
subdominant terms in Eq.~(\ref{eq:we}) therefore become
non-negligible when $q_G > \beta/L$.  Thus, the
$\beta$-independent expression (\ref{Geva}) only holds for
 low enough transport and gate voltages, such that $q_G <
\beta/L$.  In this regime our result agrees with the one obtained
by Schomerus for $\beta
 = 1$ \cite{Schomerus}.

To analyze the intermediate regime $\beta \ll q_G L \ll 1$, we keep
for $q_y < 0$ the leading term $\beta^{-1} \exp(-ik_sa_s + \eta_x L -
\zeta)$ in the denominator of Eq.~(\ref{eq:we}). This term only
dominates provided $\eta_x$ is not too small, such that $\exp (\eta_x
L) \gg 2\eta_x/q_G$. The solution of this transcendental equation, which
in the leading order becomes $\eta_x = - L^{-1} \ln (q_G L/2)$, provides
the cut-off in the integral over $\eta_x$ in the expression for the
conductance. Explicitly, we have
\begin{displaymath}
\frac{G_ev}{G_Q} = \frac{16W\beta^2}{\pi L (q_GL)^2} \ln^2 (q_GL)
\sin^2 k_sa_s \ .
\end{displaymath}
Thus, away from zero energy, the the contribution of
evanescent modes rapidly vanishes, and conductance becomes
proportional to $\beta^2$.

For yet greater gate voltages, $q_G L\gg 1$, the conductance is always
dominated by the states with $\eta_x L\lesssim 1$.  From $q_y^2 -
\eta_x^2 = q_G^2$, we therefore find that at any rate $\vert q_y \vert
\gg \vert \eta_x \vert$, and thus $\zeta = \eta_x/q_G \ll \eta_x
L$. This means that we can disregard $\zeta$ in the denominator of
Eq. (\ref{eq:we}), and in the numerator we replace $\sinh \zeta$ with
$\eta_x /q_G$. Writing $dk_y \approx \eta_x
d\eta_x/q_G$, we find that the integrand in the expression for the
conductance contains the third power of $\eta_x$ multiplied with
$e^{-\eta_x L}$. Consequently, we obtain
\begin{equation} \label{highvoltages}
\frac {G_{ev}}{G_Q} = \frac{12\zeta(3) W}{\pi L (q_G
L)^3}\frac{\sin^2k_sa_s}{|\beta^{-1}\,e^{-ik_sa_s} +
\beta\,e^{ik_sa_s} - 2|^2}.
\end{equation}
Here, $\zeta(n)$ is the Zeta
function, with $\zeta(3) \approx 1.2021$. 

The results for the conductance in different regimes are shown in
the Table.

\begin{table}
\begin{tabular}{lccc}
& & Propagating & Evanescent \\
\hline
$\beta  = 1$ & $q_G \ll 1/L$ & $q_G$ & $L^{-1}$ \\
 & $q_G \gg 1/L$ & $q_G$ & $L^{-1}(q_GL)^{-3}$\\
 \hline
$\beta \ll 1$ & $q_G \ll \beta/L$ & $q_G \beta^2$  & $L^{-1}$ \\
& $q_G \ll 1/L$ & $q_G \beta^2$ & $\beta^2 L^{-1}(q_GL)^{-2} \ln^2
 (q_GL)$ \\ 
& $q_G \gg 1/L$ & $q_G \beta$ & $\beta^2 L^{-1}(q_GL)^{-3}$
\end{tabular}
\caption{Functional dependence on the parameters $q_G$, $L$, and
$\beta = t'^2/(t_st_g)$ of contributions to the conductance from
propagating and evanescent modes in different transport regimes.}
\end{table}

\section{conclusions}
We constructed the wave-function matching conditions at the zigzag
interfaces between square (N) and graphene (G) lattices, and
determined transport properties of the NGN structure,
concentrating on the regimes of ``ideal'' interface ($t'^2 = t_at_s$)
and highly resistive interface ($t'^2 \ll t_at_s$). In accordance
with earlier predictions, at the Dirac point the conductance is
dominated by the evanescent modes and scales inversely
proportionally with the length of the contact $L$. However, the
situation changes qualitatively as soon as one departs from the
Dirac point -- for instance, by changing the electron
concentration via the gate voltage $V_G$. The propagating modes
start to contribute to the conductance.  For small $V_G$, such
that $q_G L \ll 1$, their contribution is length-independent; in
particular we find that for ideal interfaces the transmission
equals one independently of the angle of incidence. Further yet
from the Dirac point, $q_G L \gg 1$, the conductance is determined
by the propagating modes, whereas the evanescent modes'
contribution decays as $L^{-4}$. In particular, for $q_GL \gg 1$ and
$t'^2 \ll t_st_g$ we found a 
regime similar to resonant tunneling in double barrier structures. As
a consequence, the propagating modes' contribution of a long junction
in this case is greater than the one for a short junction. 

We found that the contribution of evanescent modes at the Dirac point
does not depend on $\beta$, but diminishes as we apply the gate
voltage. In the regime $q_G L \gg 1$ this contribution is suppressed.

Besides looking at the low transport bias regime considered above,
it also may be interesting to study the non-linear $I$-$V_t$
characteristics.  Our results suggest that with increasing
transport voltage $V_t$, the contribution of the evanescent modes
to the current saturates with voltage beyond $V_t
> t_g a_g/L$, while the propagating modes contribution increases
as $V_t^2$. However, the quantitative discussion of non-linear
regime is problematic, since the result would essentially depend
on the potential distribution in contacts and over the graphene
sheet. Investigation of this potential distribution would require
solution of Poisson equation coupled to the equation for the
particle density, and goes beyond the scope of this Article.

The zigzag interface considered in this Article is the simplest
case of a contact: The periods of the lattices match, $a_s =
\sqrt{3} a_g$, and the momentum component $k_y$ along the
interface is conserved.  In real experimental situations both of
these conditions will be difficult to realize: The interfaces are
disordered, and the lattice periods may be incommensurate. This
Article illustrates an importance of the interface contribution to
the transport and provides the basis for future research in this
direction.

We thank S. Trugman and H. Schomerus for useful discussions.  The
authors acknowledge Aspen 
Center for Physics, where this research was initiated. This work was supported
in part by US DOE.

{\em Note added}. At first glance, the result (\ref{highvoltages}) disagrees
with Ref. \onlinecite{Schomerus}, which finds $G_{ev} \propto L^{-1}$
for all gate voltages.However, the chemical potentials in
Ref. \onlinecite{Schomerus} were arranged in a different way than we
have done it above: The chamical potential of graphene sheet is fixed
to the Dirac point, whereas chamical potentials in the normal metal
are varied. Recently, a study of non-linear transport by Robinson and
Schomerus has been made available\cite{Robinson}. Both our above
results and the results of Ref. \onlinecite{Schomerus} follow in
appropriate limiting cases.

\end{document}